\documentclass[reprint,aps,prl,showpacs,preprintnumbers,amsmath,amssymb,superscriptaddress]{revtex4-1}

\usepackage[pdftex]{graphicx}
\DeclareGraphicsExtensions{.pdf}
\usepackage{dcolumn}
\usepackage{bm}
\usepackage{wasysym}

\begin{document}

\title{Momentum dependent ultrafast electron dynamics in antiferromagnetic EuFe$_2$As$_2$}

\author{L. Rettig}
\affiliation{Fachbereich Physik, Freie Universit\"at Berlin, Arnimallee 14, D-14195 Berlin, Germany}
\affiliation{Fakult\"at f\"ur Physik, Universit\"at Duisburg-Essen, Lotharstr. 1, D-47048 Duisburg, Germany}
\author{R. Cort\'es}
\affiliation{Fachbereich Physik, Freie Universit\"at Berlin, Arnimallee 14, D-14195 Berlin, Germany}
\affiliation{Abt. Physikalische Chemie, Fritz-Haber-Institut d. MPG, Faradayweg 4-6, D-14195 Berlin, Germany}
\author{S. Thirupathaiah}
\affiliation{Helmholtz-Zentrum Berlin, Albert-Einstein-Stra{\ss}e 15, D-12489 Berlin, Germany}
\author{P. Gegenwart}
\author{H.S. Jeevan}
\affiliation{I. Physik. Institut, Georg-August-Universit\"at G\"ottingen, D-37077 G\"ottingen, Germany}
\author{M. Wolf}
\affiliation{Fachbereich Physik, Freie Universit\"at Berlin, Arnimallee 14, D-14195 Berlin, Germany}
\affiliation{Abt. Physikalische Chemie, Fritz-Haber-Institut d. MPG, Faradayweg 4-6, D-14195 Berlin, Germany}
\author{J. Fink}
\affiliation{Helmholtz-Zentrum Berlin, Albert-Einstein-Stra{\ss}e 15, D-12489 Berlin, Germany}
\affiliation{Leibniz-Institute for Solid State and Materials Research Dresden, P.O.Box 270116, D-01171 Dresden, Germany}
\author{U. Bovensiepen}
\email[Corresponding author: ]{uwe.bovensiepen@uni-due.de}
\affiliation{Fakult\"at f\"ur Physik, Universit\"at Duisburg-Essen, Lotharstr. 1, D-47048 Duisburg, Germany}
\affiliation{Fachbereich Physik, Freie Universit\"at Berlin, Arnimallee 14, D-14195 Berlin, Germany}

\date{\today}

\begin{abstract}
Employing the momentum-sensitivity of time- and angle-resolved photoemission spectroscopy we demonstrate the analysis of ultrafast single- and many-particle dynamics in antiferromagnetic EuFe$_2$As$_2$. Their separation is based on a temperature-dependent difference of photo-excited hole and
electron relaxation times probing the single particle band and the spin density wave gap, respectively. Reformation of the magnetic order occurs at $800\,\textrm{fs}$, which is four times slower compared to electron-phonon equilibration due to a smaller spin-dependent relaxation phase space. 

\end{abstract}

\pacs{78.47.J-, 74.70.Xa, 74.25.Jb, 75.50.Ee}

\maketitle
Fe based high-$T_c$ superconductors (HTCs) exhibit, similar to the cuprate HTCs, antiferromagnetic (AFM) order in proximity to the superconducting (SC) state~\cite{Mazin2010}. The suppression of AFM ordering by doping or pressure leads to the emergence of SC. Thus, understanding the role of the AFM ground state and the coupling between low energy excitations like spin fluctuations and lattice vibrations may be an important step towards the understanding of SC in the HTCs.

One promising approach to study such interactions is to analyze the excited state of optically excited materials and its subsequent relaxation to quantify the respective coupling strengths~\cite{Brorson1990}. While in typical metals this involves electron-electron (e-e) and electron-phonon (e-ph) scattering rates~\cite{DelFatti2000, Chulkov2006}, in more complex materials with ordered states emerging in competition with thermal fluctuations, e. g. SC, charge density wave (CDW), or magnetically ordered materials, excitations specific to the ordered nature become essential~\cite{Schafer2010}. 

Unlike in ferromagnetic metals, experiments on AFM materials are hindered by absence of a net magnetization and time-resolved non-linear optical or THz spectroscopy was employed to investigate excitations of the AFM ordered state~\cite{Satoh2010, Kampfrath2011}. Metallic antiferromagnets like Fe pnictide parent compounds now offer the opportunity to analyze the interaction of low energy electrons with spin fluctuations in an antiferromagnet and to probe transient changes of the electronic band structure intimately connected to AFM order. 

\begin{figure}[tb]
\includegraphics[width=8.6cm]{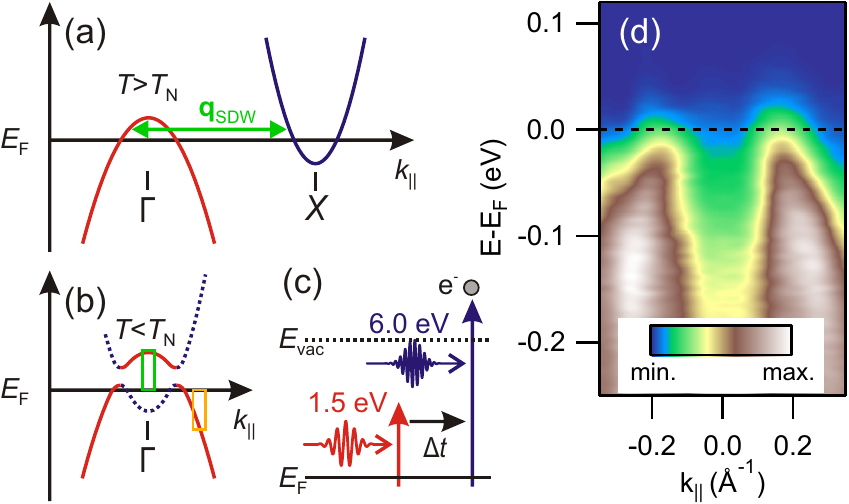}  
\caption{
\label{fig:fig1}
(color online) (a) Simplified electronic band structure of Fe pnictides along $\Gamma$-$\textit{X}$ in the paramagnetic state. (b) Below $T_{\textrm{N}}$, electron-like bands are folded back to $\Gamma$, hybridize with hole-like bands and open SDW energy gaps near $E_\mathrm{F}$. Boxes indicate the integration used in Fig.~\ref{fig:fig3}. (c) Sketch of the pump-probe experiment. (d) Laser ARPES intensity of EuFe$_2$As$_2$ taken at $T=30\,\textrm{K}$ without optical excitation.
}
\end{figure}

In Fe pnictides, the electronic bands forming the Fermi surface (FS) and determining the low-energy excitations consist of hole pockets at the center of the Brillouin zone (BZ) ($\Gamma$-point) and electron pockets at the zone corner ($\textit{X}$-point, see Fig.~\ref{fig:fig1}(a))~\cite{Mazin2010, Singh2008,Liu2008}. In undoped and weakly doped compounds, the strong nesting of hole and electron Fermi surfaces along $\textbf{q}_{\textrm{SDW}}=(\pi,\pi)$ leads at low temperatures to AFM ordering of Fe magnetic moments and the formation of a spin density wave (SDW), which is accompanied or preceded by a structural transition from te\-tra\-go\-nal to orthorhombic structure~\cite{Cruz2008,Huang2008}. Experiments demonstrated strong coupling between magnetic and structural transitions~\cite{Lester2009,Koo2010} and Ising nematic order is considered to induce orbital anisotropy~\cite{Fernandes2011,Zhang2011a}. Below the N\'{e}el transition temperature $T_{\textrm{N}}$, this magnetic ordering leads to backfolding of electron bands to $\Gamma$, where they hybridize with the hole bands and modify the FS near $\Gamma$ by a SDW energy gap, see Fig.~\ref{fig:fig1}(b)~\cite{Eremin2010}. Such partial energy gaps are thus a direct imprint of AFM ordering on the electronic band structure and their size correlates with the AFM order parameter~\cite{Yi2009,deJong2010}. 

Angle-resolved photoemission spectroscopy (ARPES) is a versatile tool to study the electronic band structure in the HTCs~\cite{Damascelli03} and has shed some light on changes in band structure in the AFM phase of FeAs systems~\cite{Yi2009, deJong2010}. Investigations of femtosecond (fs) dynamics with optical pump-probe spectroscopy~\cite{Stojchevska2010, Mertelj2010, Chia2010} concluded on enhanced interband scattering and relaxation bottlenecks across energy gaps in the AFM phase. Combining the advantages of both techniques, fs time-resolved ARPES (trARPES) allows energy- and momentum-resolved investigation of excited state dynamics in solids, e.g. Mott insulators~\cite{Perfetti2006, Petersen2011}, CDW compounds~\cite{Schmitt2008, Rohwer2011}, and cuprate HTCs~\cite{Perfetti2007, Graf2011, Cortes2011}. However, especially in complex materials the energy redistribution within the excited system and the dynamics linked to its cooperative character are often hard to disentangle. Here, trARPES can benefit from its momentum resolution and might help to separate these contributions, which is crucial for a better understanding of low energy excitations in such materials.

In this letter we report on the momentum resolved ultrafast, laser-excited electron dynamics of antiferromagnetic EuFe$_2$As$_2$ (with $T_{\textrm{N}}\approx190\,\textrm{K}$)~\cite{Jeevan2008} above and below $T_{\textrm{N}}$ as a function of temperature. We observe different relaxation dynamics of excited electrons and holes around the $\Gamma$-point which we can clearly associate with the AFM phase. We further demonstrate the separation of the excess energy dynamics, determined by the transient quasiparticle distribution, and the transient cooperative effects by analyzing the dynamics of the SDW energy gap. Simultaneously, we observe a transition to a transient paramagnetic phase on an ultrafast timescale.

We employ $55\,\textrm{fs}$ laser pulses at $\textrm{h}\nu_1=1.5\,\textrm{eV}$ with an absorbed fluence of $F=280\,\mu\textrm{J}/{\textrm{cm}}^{2}$ for optical excitation in a pump-probe experiment. A time-delayed ultraviolet (h$\nu_2=6.0\,\textrm{eV}$, $80\,\textrm{fs}$) pulse monitors the energy and momentum dependent single-particle spectral function as a function of time delay by ARPES as sketched in Fig.~\ref{fig:fig1}(c). The energy resolution of typically $50\,\textrm{meV}$ is determined by the time-of-flight spectrometer and the bandwidth of the probe pulses. The  momentum resolution is $0.05\,\textrm{\AA}^{-1}$ and the temporal resolution is $<100\,\textrm{fs}$~\cite{Schmitt2008}. Single crystals of EuFe$_2$As$_2$ grown by the Bridgman method~\cite{Jeevan2008} were cleaved \textit{in situ} in ultrahigh vacuum ($p<4\times10^{-11}\,\textrm{mbar}$) prior to the measurements. 

\begin{figure}[tb]
\includegraphics[width=8.6cm]{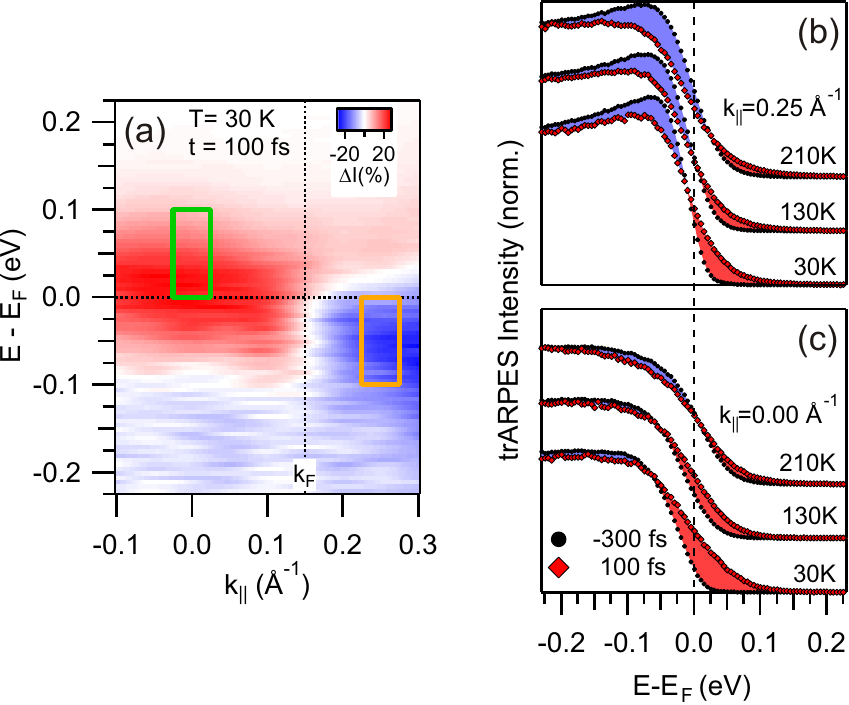}  
\caption{
\label{fig:fig2}
(color online) (a) Pump-induced change of spectral weight $\Delta I$ in a false color map for $t=100\,\textrm{fs}$. Blue color marks depletion of spectral weight (hole excitations), while red marks increased $\Delta I$ (electron excitations). Boxes represent integration areas for $\Delta I$ used in Fig.~\ref{fig:fig3}. (b) Energy distribution curves (EDCs) for $k_{\parallel}=0.25\,\textrm{\AA}^{-1}>k_\mathrm{F}$ before (black) and $t=100\,\textrm{fs}$ after laser excitation (red) for various temperatures. Increase and depletion of spectral weight are marked by red and blue areas, respectively. Spectra are vertically offset for clarity. (c) EDCs for $k_{\parallel}=0.00\,\textrm{\AA}^{-1}$ ($\Gamma$).
}
\end{figure}

The laser ARPES intensity of EuFe$_2$As$_2$ at h$\nu$=6.0~eV without excitation is shown in Fig.~\ref{fig:fig1}(d) as a function of energy $E-E_\textrm{F}$ and momentum $k_\parallel$. The prominent dispersion of the hole pocket around the $\Gamma$-point compares well with data taken at higher photon energies~\cite{deJong2010}. Optical excitation causes considerable changes to the spectral weight $I(E,k)$ as shown in a false-color plot of the pump induced change $\Delta I(E,k,t) = I(E,k,t) - I_0(E,k)$ for $t=100\,\textrm{fs}$ after excitation in Fig.~\ref{fig:fig2}(a). Two distinct types of response are found: near the $\Gamma$-point within the hole pocket, an increase in spectral weight is observed (red), whereas we find a decrease for occupied states outside the hole pocket (blue). We identify this increase (decrease) as hot electrons (holes) which originate from secondary electron-hole (e-h) pair excitations, created by e-e scattering of primarily excited electrons. This leads to the excitation of electrons from occupied states outside the hole pocket into unoccupied states inside the pocket.

In order to get more insight into the origin of $\Delta I$, Fig.~\ref{fig:fig2}(b) and (c) compare energy distribution curves (EDCs) before and after excitation at representative momenta for electron and hole excitations at $\Gamma$ and at $k_{\parallel}=0.25\,\textrm{\AA}^{-1}>k_\mathrm{F}$, respectively, and for various temperatures across the AFM transition. For states outside the hole pocket (Fig.~\ref{fig:fig2}(b)) we see a transient excitation of e-h pairs symmetric to $E_\mathrm{F}$, as expected for a single band metal~\cite{Fann1992, Lisowski2004}. We find that this behavior is robust upon cooling below $T_\mathrm{N}$, which is very reasonable as the magnetic order mainly affects states near $\Gamma$. 

At $\Gamma$ the situation is clearly different (see Fig.~\ref{fig:fig2}(c)). Above $T_\mathrm{N}$, the system behaves like a metal with e-h excitations symmetric to $E_\mathrm{F}$, similar to $k_{\parallel}=0.25\,\textrm{\AA}^{-1}$. With lowering $T<T_\mathrm{N}$, however, we find before excitation a shift of the leading edge to lower energies, consistent with the opening of a SDW energy gap (see Fig.~\ref{fig:fig1}(b)). Upon photoexcitation, this gap closes and is filled by electrons, evidenced by the shift of the leading edge and the strong increase of spectral weight at $E_\mathrm{F}$ ((Fig.~\ref{fig:fig2}(c)), red area). As this redistribution is only observed for $T<T_\mathrm{N}$, we conclude that it is closely linked to AFM order. This suggests that analyzing the dynamics of the electron excitations at $\Gamma$ enables us to investigate the collective dynamics of the transient AFM order. Simultaneously, as shown below, the hole excitations at $k_{\parallel}>k_\mathrm{F}$ serve as a measure for single-particle e-ph scattering dynamics.

\begin{figure}[tb]
\includegraphics[width=8.6cm]{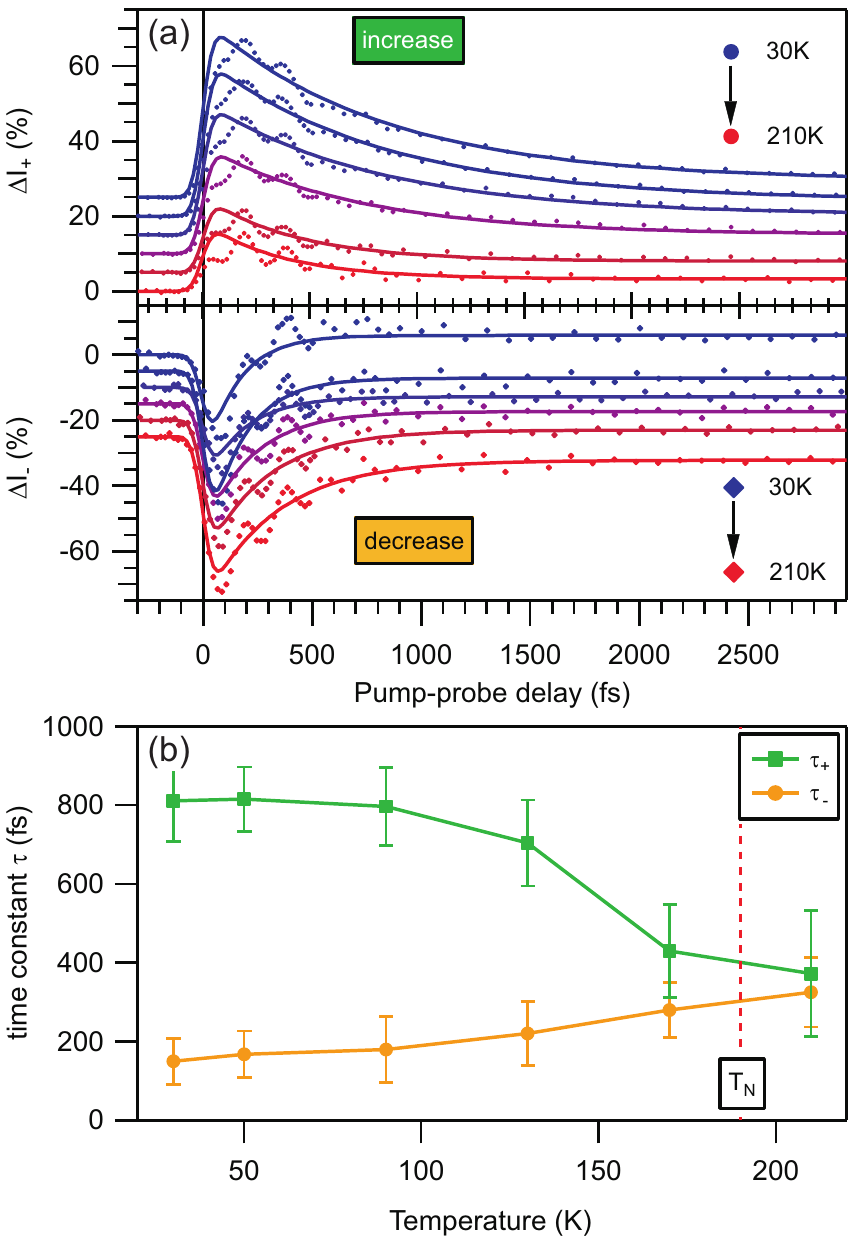}  
\caption{
\label{fig:fig3}
(color online) (a) Time-dependent spectral weight integrated over the energy-momentum intervals in Fig.~\ref{fig:fig2}(a), at $\Gamma$ (upper panel, increase) and $k_{||}=0.25\,\textrm{\AA}^{-1}$ (lower panel, decrease) for various temperatures. Data are vertically offset for clarity. Solid lines are exponential fits. (b) Relaxation time constants $\tau_{+}$ and $\tau_{-}$ as function of temperature obtained from the fits. Error bars represent $99.7\%$ confidence intervals.
}
\end{figure}

To analyze the time evolution of the momentum-dependent e-h distribution, we integrate the data over representative intervals for electron and hole excitations, depicted in Fig.~\ref{fig:fig2}(a). Thereby we determine the temporal evolution of pump-induced spectral weight for hole excitations $\Delta  I_{-}$ (yellow box) and for electron excitations $\Delta I_{+}$ (green box), depicted in Fig.~\ref{fig:fig3}(a) for various $T$. $\Delta I_{+}$ ($\Delta I_{-}$) shows an ultrafast increase (decrease) within the pump-pulse duration and a relaxation in $1-2\,\textrm{ps}$. We notice a strong variation with temperature in the pump-induced response for $\Delta I_{+}$, while $\Delta I_{-}$ shows only minor variations. 
A closer look on the data reveals a transient oscillation of $\Delta I$ superimposed on the decay with a period of $\sim200\,\textrm{fs}$, originating from the coherently excited A$_{1g}$ phonon mode~\cite{Mansart2009, Mertelj2010}, which changes the Fe-As distance perpendicular to the Fe-As planes and will be subject to a future publication. 

Next, $\Delta I_{+,-}$ are fitted with exponential decay functions, $\Delta$$I_{+,-}(t)=A \exp(-t/\tau_{+,-})+B$, convoluted with the temporal pump-probe envelope. Here, $A$ is the excitation amplitude, $\tau$ the relaxation time constant and $B$ accounts for lattice heating after e-ph thermalization~\cite{Lisowski2004}. Fig.~\ref{fig:fig3}(b) shows the obtained relaxation constants $\tau_{+,-}$ as a function of $T$. For $T>T_\mathrm{N}$, both electron and hole populations relax on the same timescale of $\sim400\,\textrm{fs}$, governed by e-ph scattering (see below). By lowering $T$ below $T_\mathrm{N}$, however, we find a strong difference in $\tau$ between electron and hole relaxation. While the relaxation of excited holes shows a slight acceleration for lower $T$, the electron relaxation time at $\Gamma$ strongly increases. At $T=30\,\textrm{K}$, holes decay more than four times faster ($\tau_{-}<200\,\textrm{fs}$) than electrons, which relax with $\tau_{+}\sim800\,\textrm{fs}$.

We now discuss these peculiar temperature dependent relaxation times $\tau_{+}$ and $\tau_{-}$. Hole excitations outside the hole pocket are barely influenced by the presence of AFM ordering. Their decay ($\tau_{-}$) is governed by the energy transfer to the lattice through e-ph relaxation. This is supported by the slower relaxation with increasing $T$, as more and more phonon modes are occupied and thus reduce the phase space for e-ph scattering. In the limit of a bad metal, $\tau_{-}$ is expected to depend linearly on $T$~\cite{Kabanov2008, Stojchevska2010}, which allows determination of the second moment of the Eliashberg e-ph coupling function, $\lambda<\omega^2>$. Indeed, the value of $\lambda<\omega^2>=90\pm40\,\textrm{meV}^2$, which we determine for $T>100\,\textrm{K}$ from our data following Ref.~\cite{Stojchevska2010} is in agreement with reports for SrFe$_2$As$_2$~\cite{Stojchevska2010} and SmAsFeO~\cite{Mertelj2010}, further supporting our assignment.

In contrast, electrons at $\Gamma$ are strongly influenced by spin-dependent interactions in the AFM phase and thus the observed dynamics is linked to the transient change of the AFM order. Remarkably, the observed temperature dependence of $\tau_{+}$ strikingly resembles the temperature dependence of the SDW order parameter~\cite{Herrero-Martin2009} and saturates at $\tau_{+}=800\pm100\,\textrm{fs}$ below $100\,\textrm{K}$, similar to the relaxation found in antiferromagnetic SmAsFeO~\cite{Mertelj2010}. Qualitatively, we can understand the slower recovery dynamics of the AFM order ($\tau_{+}$) compared to the e-ph relaxation ($\tau_{-}$) by a reduced phase space available for e-e scattering across the energy gap and a spin-spin scattering bottleneck. With increasing temperature, thermal fluctuations of the spin order provide additional relaxation channels and thus decrease $\tau$. Above $T_\mathrm{N}$ with the breakdown of spin order, the relaxation bottleneck is absent and the decay is governed by e-ph scattering on the same timescale as the hole excitations. 

\begin{figure}[tb]
\includegraphics[width=8.6cm]{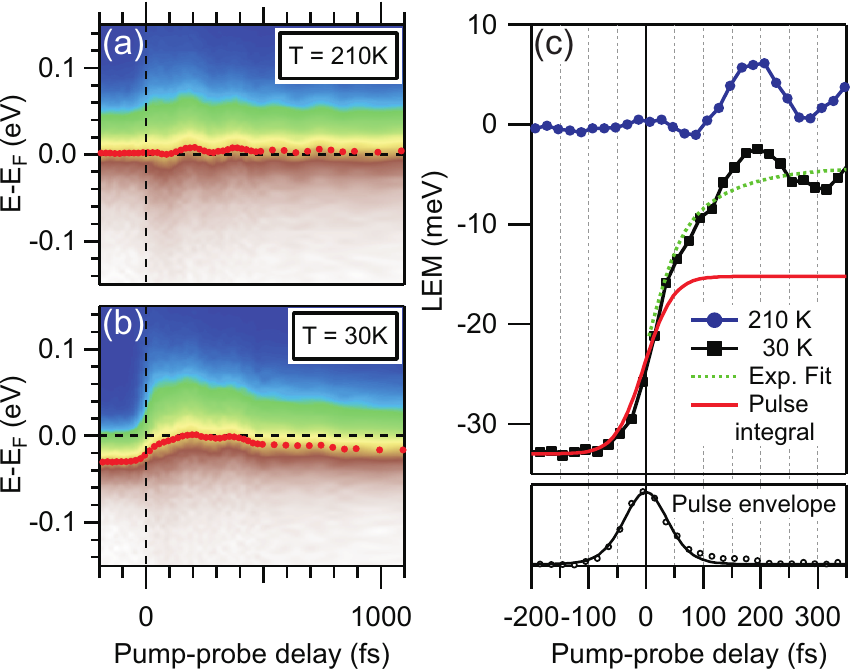}  
\caption{
\label{fig:fig4}
(color online) trARPES intensity in a false color representation at $\Gamma$ for $T=210\,\textrm{K}$ (a) and $T=30\,\textrm{K}$ (b), as function of energy and pump-probe delay. Red dots mark the position of the leading edge midpoint (LEM) of the spectra. (c) Upper panel: LEM position around zero pump-probe delay. The pulse integral scaled to the data is shown as red solid line, and the green dashed line is an exponential fit (see text). Lower panel: Pulse envelope obtained from electrons at $E-E_\mathrm{F}>1\,\textrm{eV}$.
}
\end{figure}

Finally, we address the photo-induced perturbation of spin order. Fig.~\ref{fig:fig4}(a) and (b) show trARPES intensities at $\Gamma$ as function of energy and pump-probe delay for $T=210\,\textrm{K}$ and $T=30\,\textrm{K}$, respectively. To investigate the 
response of the SDW gap to the optical excitation, the leading edge midpoint (LEM) is shown in Fig.~\ref{fig:fig4}(c) for every delay point. At $T=210\,\textrm{K}>T_\mathrm{N}$, the LEM matches the Fermi level $E_\mathrm{F}$ and shows no significant modification around time zero. After $\sim100\,\textrm{fs}$, the coherent A$_{1g}$ mode starts modulating the LEM. 

In the AFM phase at $T=30\,\textrm{K}$, the LEM is shifted by $\sim-30\,\textrm{meV}$ before excitation, which is compatible with values of the SDW gap of $2\Delta\sim90\,\textrm{meV}$ measured by optical spectroscopy~\cite{Wu2009}. Upon excitation, we find a strong upshift towards the high-temperature value, i.e. a closing of this gap. Thus, we interpret this shift as a measure of the transient AFM order parameter, which facilitates a quantitative analysis of its dynamics. We identify two regimes: First, until $t=50\,\textrm{fs}$, we find a very fast shift of the LEM, which follows the pump-probe pulse integral (solid line). Later, the LEM continues to shift on a timescale of $\tau\sim100\,\textrm{fs}$, as indicated by the dashed line. It peaks at $200\,\textrm{fs}$ when it has nearly reached the high-temperature value, which suggests a major loss of AFM order. The LEM likewise starts to become modulated by the A$_{1g}$ mode and subsequently decays towards the initial value, governed by the recovery of the AFM state.

In collective lattice dynamics, e.g. in CDW materials, the material's response is retarded due to the lattice inertia~\cite{Schmitt2008}. In contrast, purely electronic phenomena like Mott~\cite{Perfetti2006, Petersen2011,Wall2011} and SDW transitions happen on faster timescales. Hence, the observation of two timescales here suggests the ultrafast decoupling of the coupled AFM and structural transitions. First, the initial fast shift of the LEM shows the ultrafast reduction of long-range AFM ordering, whereas local order, e.g. Ising nematic like, persists longer and is destroyed by the structural reorientation on the timescale of $100\,\mathrm{fs}$. Future fluence dependent studies might shed light on the coupled or competing character of the underlying fluctuations.

In conclusion, using trARPES we investigated the fs electron and hole dynamics near the $\Gamma$-Point in EuFe$_2$As$_2$ in the antiferromagnetic and paramagnetic phase. We find a strong difference in electron and hole dynamics induced by the AFM order, which allows us to separate its dynamics from single-particle dynamics like e-ph scattering. In the AFM phase, the restricted phase space for relaxation of electrons at $\Gamma$ and a spin-relaxation bottleneck leads to a slow recovery of AFM ordering with $\tau=800\,\textrm{fs}$, while e-ph scattering is more than four times faster. Analysis of the leading edge midpoint of trARPES spectra suggests that the initial loss of spin order is dominated by purely electronic processes, followed by a structural reorientation on a slower timescale of $\tau\sim100\,\textrm{fs}$.

\begin{acknowledgments}
We acknowledge fruitful discussions with I. Eremin and funding by the DFG through BO 1823/2 and SPP 1458, as well as the AvH Foundation.
\end{acknowledgments}

%

\end{document}